\documentclass[%
 twocolumn,
superscriptaddress,
 amsmath,amssymb,
 aps,
floatfix,
]{revtex4}

\usepackage{graphicx}
\usepackage{dcolumn}
\usepackage{mathrsfs}
\usepackage{bm}
\usepackage{xcolor}

\newcommand{\Pa}[0]{{\mathrm{P}}}
\newcommand{\co}[0]{{\mathrm{co}}}
\newcommand{\V}[0]{{\mathrm{V}}}
\newcommand{\crit}[0]{{\mathrm{c}}}
\newcommand{\Rel}[0]{{\mathrm{Rel}}}

\newcommand{\beginsupplement}{%
        \setcounter{table}{0}
        \renewcommand{\thetable}{S\arabic{table}}%
        \setcounter{figure}{0}
        \renewcommand{\thefigure}{S\arabic{figure}}%
     }

\begin{document}

\preprint{APS/123-QED}

\title{Dynamics of Spontaneous Wrapping of Microparticles by Floppy Lipid Membranes
}

\author{Hendrik T. Spanke}
\affiliation{ETH Z\"{u}rich}%
\author{Jaime Agudo-Canalejo}
\affiliation{Max Planck Institute for Dynamics and Self-Organization (MPIDS)}
\author{Daniel Tran}
\affiliation{ETH Z\"{u}rich}%
\author{ Robert W. Style}%
\affiliation{ETH Z\"{u}rich}%
\author{Eric R. Dufresne}
\email{eric.dufresne@mat.ethz.ch}%
\affiliation{ETH Z\"{u}rich}%

\date{\today}

\begin{abstract}

Lipid membranes form the barrier between the inside and outside of cells and many of their subcompartments.
As such, they bind to a wide variety of nano- and micrometer sized objects and, in the presence of strong adhesive forces, strongly deform and envelop particles.
This wrapping plays a key role in many healthy and disease-related processes.
So far, little work has focused on the dynamics of the wrapping process.
Here, using a model system of micron-sized colloidal particles and giant unilamellar lipid vesicles with tunable adhesive forces, we measure the velocity of the particle during its wrapping process as well as the forces exerted on it by the lipid membrane.
Dissipation at the contact line appears to be the main factor determining the wrapping velocity and time to wrap an object.

\end{abstract}

\maketitle

Lipid membranes frequently come in contact with nano- and micro-objects. 
This is essential in many biological processes.
Examples range from the disease-related entry of viruses and bacteria into cells \cite{Flannagan2012,Dasgupta2014} to healthy docking and priming during vesicular trafficking \cite{Sudhof2013}.
A large body of work has studied adhesive particle-membrane interactions \cite{Lipowsky1998,Bahrami2014,Dasgupta2014a,Reynwar2007,Bahrami2012,Raatz2014, Xiong2017,Koltover1999,ruiz2012,Saric2012, Saric2013, Vahid2017}.
However, only a few studies have focused on adhesion dynamics \cite{Mirigian2013,Dietrich1997}.

In theory, the interplay of a simple membrane with  particles should be governed by only a few physical parameters.
The membrane resists bending through its bending rigidity, $\kappa_\mathrm{b}$, and stretching through its membrane tension, $\sigma$.
The particle-membrane adhesion is characterized by the adhesion energy per unit area, $\omega$.
Above a critical adhesion energy, $\omega_\crit$, a membrane will spontaneously wrap a particle coming into contact with it.
For flat, tensionless membranes, 
$\omega_\crit=2\kappa_\mathrm{b}/R_\Pa^2$, where $R_\Pa$ is the particle radius \cite{Spanke2020,Agudo-Canalejo2015}.

Here, we experimentally investigate the spontaneous wrapping of micron-sized particles by giant unilamellar vesicles (GUVs) in the biologically relevant limit of low membrane tension and weak reversible adhesion.
We tune particle-membrane interactions using the depletion effect, and investigate how the wrapping dynamics change with increasing $\omega$.
 Comparing spontaneous wrapping of free and  optically-trapped particles, we find that dissipation at the membrane-particle contact line   controls the wrapping dynamics.

We use a recently developed model system consisting of a dispersion of micron-sized polystyrene particles ($0.54 \pm 0.02\ \mathrm{\mu m}$ and $1.04 \pm 0.02\ \mathrm{\mu m}$ in radius) and  GUVs ($9.8-24.6~\mathrm{\mu m}$ in radius) combined with a depletant \cite{Spanke2020}.
See the Supplemental Material for a histogram of GUV sizes used.
The GUVs are made by electroformation in a 280 mOsm/kg sucrose solution.
They consist of 1-palmitoyl-2-oleyl-sn-glycero-3-phosphocholine (POPC) with 1\% 1,2-dioleoyl-sn-glycero-3-phosphoethanolamine-N-(lissamine rhodamine B sulfonyl) (Rhodamine PE) \cite{Angelova1986,Angelova1992,dimova2019}.
We previously measured $\kappa_\mathrm{b}$ in this system to be $33\pm8~\mathrm{k_B T}$ \cite{Spanke2020}.

The model system allows us to tune a number of parameters.
We can vary $\omega$ by changing the concentration of polyethylene glycol (PEG) depletant.
This has a molecular weight of $10^5$ g/mol, a radius of gyration $R_g$ of about 16 nm, and an overlap concentration of 0.99 wt\% \cite{Ziebacz2011, rubinstein2003polymer}.
We used PEG concentrations between 0.39-0.67 wt\% ($\pm$ 0.016 wt\%) in the samples.
When taking a steric repulsion due to strong thermal fluctuations of the membrane into account this yields adhesion energy densities from $0.9-1.7~\mathrm{\mu J/m^2}$.
Previous studies into the dynamics of particle wrapping consider much higher adhesion energy densities of $10^{2}-10^{3}~\mathrm{\mu J/m^2}$ \cite{Dietrich1997,Mirigian2013}.
The osmolality of the outside solvent is adjusted  through the addition of glucose (approximately 270 mM) to a slightly hypertonic value of 290 mOsm/kg including 10 mM of sodium chloride, the PEG depletion agent and a small amount of Pluronic F108.
Over the course of hours, this slight osmotic imbalance drives the deflation of vesicles, leading to very low membrane tensions on the order of $10^{-10}~\mathrm{N/m}$ \cite{Spanke2020}.
With these low membrane tensions the adhesion energy driving the wrapping process is counteracted only by the bending energy of the membrane.

\begin{figure}
\includegraphics[width=0.9\columnwidth]{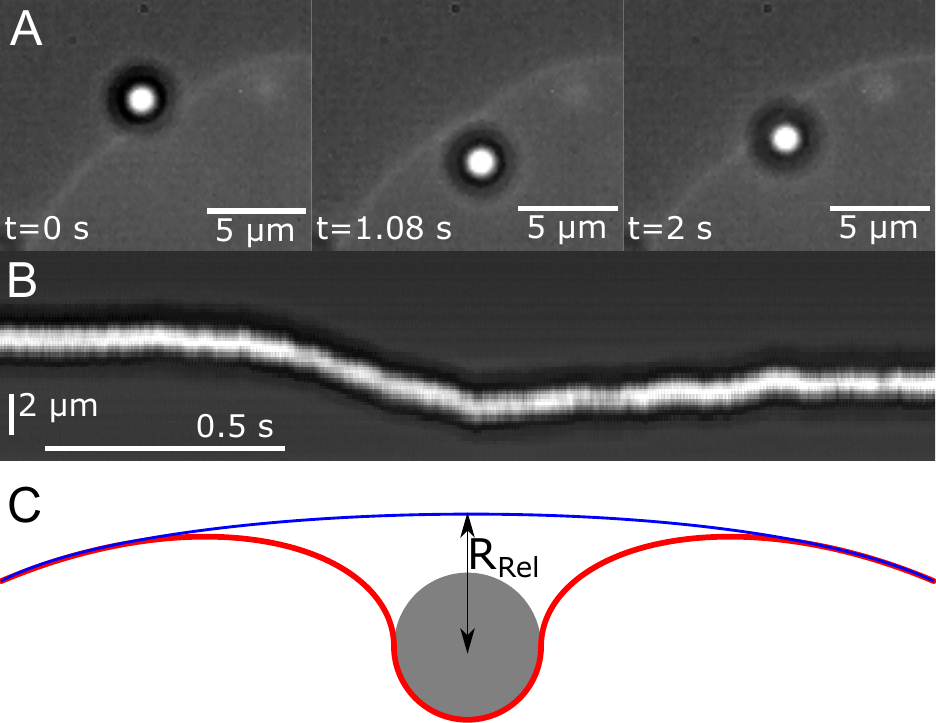}
\caption{\label{fig:RRel} \emph{Spontaneous Wrapping}.  
\emph{(A)} Combined brightfield and fluorescence microscopy images of a PS particle $1.04~\mathrm{\mu m}$ in radius being spontaneously wrapped by a POPC membrane in the presence of 0.67 wt\% PEG100K.
\emph{(B)} A kymograph showing the particle movement along an axis perpendicular to the membrane.
\emph{(C)} Schematic introducing the $R_\Rel$ coordinate. $R_\Rel$ expresses the shortest distance between the particle center and a parabola, fit to the membrane outside of the immediate deformation caused by the particle.}
\end{figure}

Above a critical adhesion energy density, $\omega_\crit$, a particle is spontaneously engulfed by the membrane, once it comes within range of the depletion interactions.
The spontaneous wrapping process is presented in Fig. \ref{fig:RRel}A.
The particle is observed to be quickly engulfed after being moved close to a GUV using an optical trap and coming into contact with the membrane.
This process can be separated into two parts.
At first, the particle moves about two particle radii in the direction of the vesicle center.
It reaches its furthest indentation after 1.08 seconds in the example shown in Fig. \ref{fig:RRel}A, before moving away from the vesicle center again and settling just underneath the membrane.
During the whole process, the particle moves  in a radial direction with respect to the GUV center.
Fig. \ref{fig:RRel}B shows a kymograph along this perpendicular direction marking the particle trajectory over time.
Wrapping of the particle is accompanied by movement of the particle as well as large-scale deformation of the membrane.
Additionally, the membrane shows large thermal fluctuations with amplitudes on the order of $\mathrm{\mu m}$, owing to the very low membrane tension in this system \cite{Spanke2020}.
For these reasons we chose to express the wrapping progress through the position of the particle relative to the membrane, $R_\Rel$, defined in Fig. \ref{fig:RRel}C. 
In experiments, the close range deformation of the membrane induced by the particle cannot be resolved, Fig. \ref{fig:RRel}A.
Outside of the immediate vicinity of the particle the membrane will return to its undisturbed equilibrium shape.
By fitting a parabola to this membrane segment, the undisturbed membrane at the particle position can be interpolated.
We define $R_\Rel$ as the shortest distance between the tracked particle position and this parabola in each frame of the acquisition.

\begin{figure}
\includegraphics[width=0.9\columnwidth]{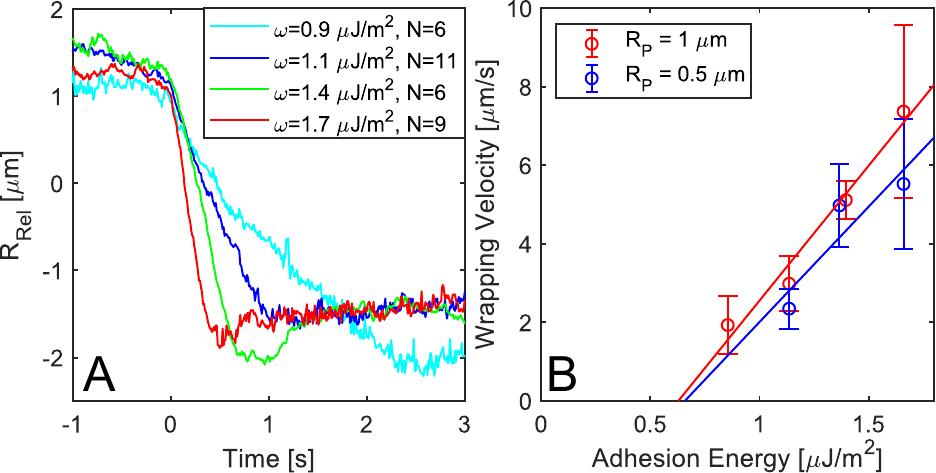}
\caption{\label{fig:Exp} \emph{Wrapping Trajectories and Velocities.}.
\emph{(A)} Averaged trajectories for a particle with a radius of $1.04~\mathrm{\mu m}$ with increasing adhesion energy densities. Each curve is the average of N individual experiments, as is indicated in the legend. 
\emph{(B)} Wrapping velocity as a function of adhesion energy density for particles with a radius of $1.04~\mathrm{\mu m}$ in red and $0.54~\mathrm{\mu m}$ in blue.
}
\end{figure}

We probed a number of GUVs in samples with different $\omega$.
Videos were acquired at $1000$ frames per second.
The membrane-tracking was particularly susceptible to single-frame outliers in its tracked position, due to the resolution of the camera and low-light conditions at these high frame rates.
The individual curves acquired were median-filtered over ten frames to filter out this noise.
Fig. \ref{fig:Exp}A shows $R_\Rel$ over time for four different adhesion energies ranging from $0.9$ to $1.7~\mathrm{\mu J/m^2}$ and particles with a radius of $1.04~\mathrm{\mu m}$.
Each curve is the average of $N$ individual experiments, as indicated in the legend.
Averaged $R_\Rel$-trajectories for particles with a radius of $0.54~\mathrm{\mu m}$ are shown in fig. \ref{fig:S_1mu} in the Supplement.
In these $R_\Rel$-trajectories, the slope of the initial uptake appears linear, allowing for a linear fit to extract the wrapping velocity.
Fig. \ref{fig:Exp}B shows the wrapping velocity extracted from the $R_\Rel$-trajectories for particles with a radius of $1.04~\mathrm{\mu m}$ and $0.54~\mathrm{\mu m}$.
The error bars indicate the standard deviation of the experiments.
The observed wrapping velocities vary between about $2$ and $8~\mathrm{\mu m/s}$.
The wrapping velocities are not significantly affected by the particle size.
Extrapolating the linear trend to much higher adhesion energy densities used in previous studies, we find wrapping velocities close to the reported values \cite{Dietrich1997}.
Fig. \ref{fig:Exp}B suggests a critical adhesion energy $\omega_\crit$ of about $0.65~\mathrm{\mu J/m^2}$ at which the observed wrapping velocity goes to zero and no spontaneous wrapping should take place.
From previous experiments, we expect the critical adhesion energy density to be in the range of $0.28-0.31~\mathrm{\mu J/m^2}$ and $0.97-1.04~\mathrm{\mu J/m^2}$ for particles with a radius of $1.04~\mathrm{\mu m}$ and $0.54~\mathrm{\mu m}$ respectively, depending on the GUV radius \cite{Spanke2020}.

\begin{figure}
\includegraphics[width=0.9\columnwidth]{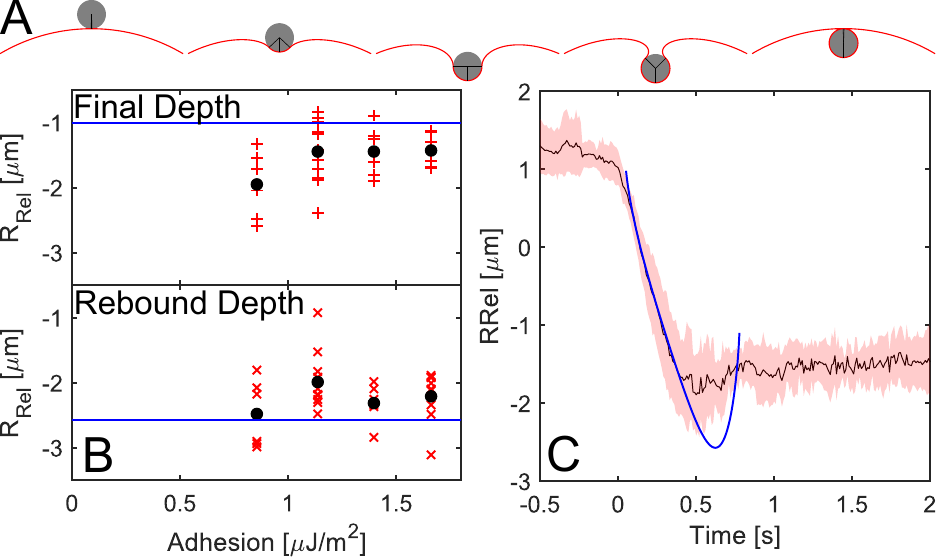}
\caption{\label{fig:Model} \emph{Quasistatic Model}.  
\emph{(A)} Quasistatic membrane shapes during the wrapping process for wrapping degrees $\mathrm{\textit{q}}$ of 0.01, 0.26, 0.5, 0.75 and 0.99 for a particle to vesicle radius ratio of $0.08$.
This ratio corresponds to a vesicle radius $R_\V$ of $13~\mu m$, a typical size observed in experiments, and a particle radius $R_\Pa$ of $1.04~\mu m$.
The membrane is kept at the same level between schematics.
\emph{(B)} The final $R_\Rel$-position at which the particle settles after wrapping and the deepest indentation during the process. The red x indicate each individual experiment and the black dot the average of those.
The blue line is the expected final position and deepest indentation of the particle from quasistatic models.
\emph{(C)} Averaged $R_\Rel$-curve for a particle with a radius of $1.04~\mathrm{\mu m}$ at an adhesion energy density of $1.7~\mathrm{\mu J/m^2}$.
The standard deviation in $R_\Rel$ at each timepoint is indicated in red.
The blue line is the modelled $R_\Rel$.
}
\end{figure}

To understand the  shape of $R_\Rel$ over time we need to consider the underlying membrane mechanics.
The membrane will conform to the particle curvature as it wraps around the particle.
A degree of wrapping, $q$, can be defined as the fraction of the particle surface area bound to the membrane \cite{Agudo-Canalejo2017}.
The distance $z$ between the bottom of the particle and the contact line is then given by  $z=2R_\Pa q$.
The equilibrium shapes of the membrane can be calculated  for different levels of wrapping.
They are calculated numerically by solving the membrane shape equations using a shooting method \cite{Seifert1991,Agudo-Canalejo2016}.
The membrane is taken to be part of a vesicle with constant area $A_V$ and variable enclosed volume, so that in the absence of the particle it forms a sphere with radius $R_V = \sqrt{A_V/4\pi}$.
The membrane shape at different stages of a quasi-static wrapping process are shown for $R_p/R_V=0.08$ in Fig. \ref{fig:Model}A.
The particle initially indents  the membrane,  then rebounds at $q \approx 0.75$.
This quasistatic model, described further in the Supplement, allows for the extraction of the same $R_\Rel$-coordinate used in the experiments.
It predicts that the particle  rebounds at  a depth $R_\Rel\approx-2.5R_\Pa$ before settling at a final depth of $R_\Rel\approx-R_\Pa$.
In experiments, however, the rebound is more shallow than expected ($R_\Rel\approx-2R_\Pa$), while the final depth is deeper than expected ($R_\Rel\approx-1.5R_\Pa$), as shown in Fig. \ref{fig:Model}B.
Additionally, the final depth shows a slight dependence on the adhesion energy, with the particle resting closer to the undeformed membrane at higher adhesion energies.   This is not expected from the model.
As shown in the Supplement, thermal fluctuations could account for part of this discrepancy, increasing the final depth to  about $R_\Rel=-1.2~\mathrm{\mu m}$ at $\omega_\crit$ (see Supplement for details).
Furthermore, the depletion interactions which bind particles to the membrane also favor adhesion of the membrane to itself.
This could drive extension of the narrow neck connecting the fully wrapped particle to the rest of the membrane.

In order to predict the time-evolution of wrapping, 
we consider the particle dynamics as a function of $z$.
Assuming overdamped dynamics, 
\begin{equation}
\dot{z} = -\mu E'(z)
\label{eq:Maindyn1}
\end{equation}
where $E$ is the energy of the membrane-particle system, described in the Supplement.
Assuming that the mobility, $\mu$,  is independent of $z$, we integrate  Eq. \ref{eq:Maindyn1}  to arrive at the blue $R_\Rel$-trajectory shown in Fig. \ref{fig:Model}C. 
Here, the model was fit to the data using the mobility as a free parameter, with reasonable agreement in the initial stages of wrapping for $\mu=4.2\times10^{5}~\mathrm{(N~s/m)}^{-1}$, see Supplement for more details on the fit.
This value for the mobility, $\mu$, is about 60 times lower than the Stokes mobility ($1/6\pi \eta R_\Pa$) drag on the same sphere in a fluid of the same viscosity, $\eta=0.0019~\mathrm{Pa~s}$

\begin{figure}
\includegraphics[width=0.9\columnwidth]{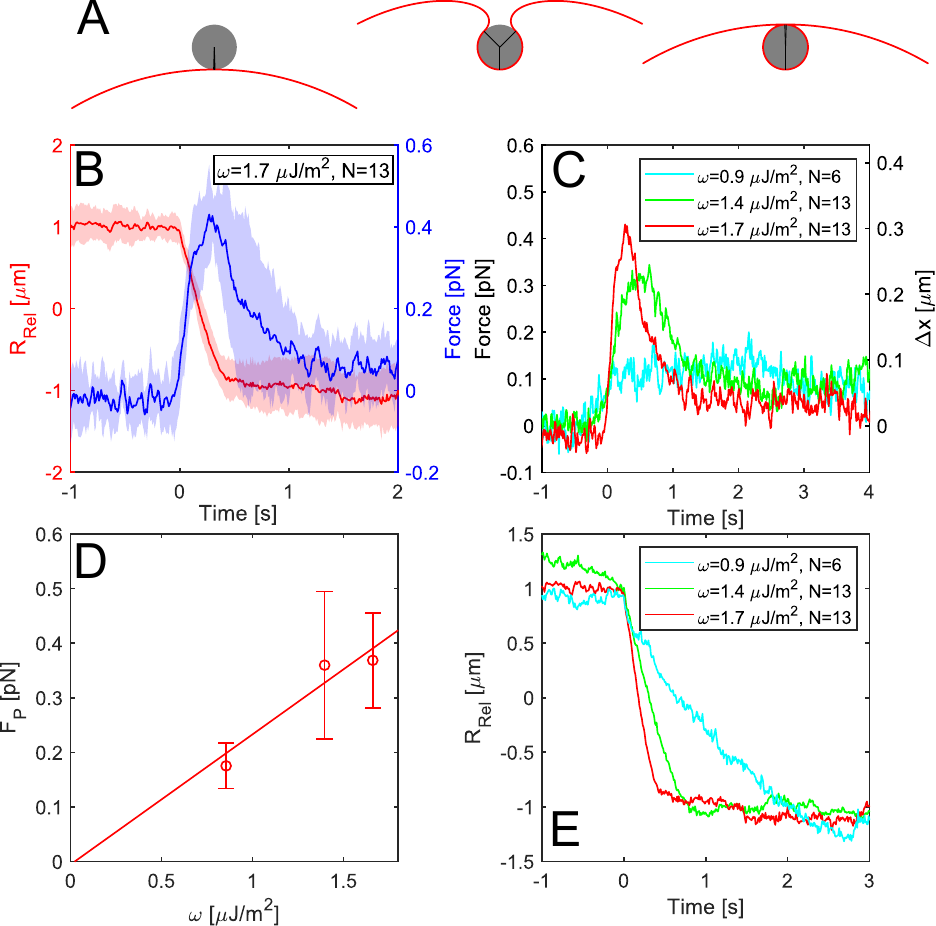}
\caption{\label{fig:Force} \emph{Force Measurement}.  
\emph{(A)} Quasistatic membrane shapes during the wrapping process for wrapping degrees $\mathrm{\textit{q}}$ of 0.01, 0.75 and 0.99 for a particle to vesicle radius ratio of $0.08$.
This ratio corresponds to a vesicle radius $R_\V$ of $13~\mu m$, a typical size observed in experiments, and a particle radius $R_\Pa$ of $1.04~\mu m$.
The particle is kept at the same level between schematics.
\emph{(B)}Averaged $R_\Rel$ and force for a particle with a radius of $1.04~\mathrm{\mu m}$ at an adhesion energy density of $1.7~\mathrm{\mu J/m^2}$.
\emph{(C)} Averaged force for increasing adhesion energy densities $\omega$.
The displacements indicated on the right vertical axis correspond to the average trap stiffness, $1.4~\mathrm{pN/\mu m}$, of all experiments.
\emph{(D)} Peak force as a function of adhesion energy density for particles with a radius of $1.04~\mathrm{\mu m}$.
\emph{(E)} Averaged $R_\Rel$-curves for increasing adhesion energy densities $\omega$. Each curve is the average of \textit{N} individual experiments, as is indicated in the legend.
}
\end{figure}

To gain more insight into possible dissipation pathways during the wrapping process, we consider the wrapping dynamics of a particle trapped in optical tweezers.
We set the trap stiffness to $1.3 - 1.7~\mathrm{pN/\mu m}$.  In this regime, the particle displacements are small but resolvable, so that we can reliably measure forces.
To completely wrap the particle, the membrane now has to deform and move past the particle as it is wrapping (Fig. \ref{fig:Force}A).
As shown in Fig. \ref{fig:Force}B, the force on the particle increases at the onset of wrapping, and reaches its peak just as the particle becomes fully wrapped ($R_\Rel=-R_\Pa$).
Afterward, the force slowly decays to zero as the vesicle relaxes to its equilibrium shape in the far-field.
The peak forces, $F_p$, increase with the adhesion energy, as shown in Fig. \ref{fig:Force}CD.
Fig. \ref{fig:Force}C shows the measured forces for three different values of $\omega$.
The peak forces are in the range of $0.1$ to $0.5~\mathrm{pN}$ and increase with $\omega$.

Interestingly, the relative motion of the particle and membrane is very similar for  free  (Fig. \ref{fig:Force}E) and trapped  (Fig. \ref{fig:Exp}A) particles. 
Specifically, the timescales of the wrapping processes very similar, lasting just over a second for the highest adhesion energy in both experiments.
This suggests that  dominant time scale depends on neither the relative motion of the particle and fluid, nor the membrane and fluid.  Instead, it is dominated by the relative motion of the particle and membrane.
Indeed, we expect significant dissipation associated with draining the fluid between the particle and membrane  as they establish  intimate contact \cite{Cantat2003,Bernard2000}.
Previous theory has quantified the size of the region near the contact line where dissipation takes place, $\ell_D=(2\kappa/\omega)^{(1/4)} h_0^{(1/2)}$ \cite{Blount2015}.
Here, $h_0$ is the distance between particle and membrane at which the two are considered to be touching.
The equilibrium separation distance between the particle and the fluctuating membrane in the system, $h_0 \approx 3.9-4.7~\mathrm{nm}$ is given by $\left[\left(2c k_\mathrm{B}T\right)/\left(\kappa_\mathrm{b} n\right)\right]^{(1/3)}$ \cite{Helfrich1984,Spanke2020}.
Therefore,  the size of the dissipation zone in our case is expected to be $\ell_D = 30-40~\mathrm{nm}$.  This is much smaller than $R_\Pa$.
Therefore, dissipation is localized to a small region near the contact line where the membrane and particle meet.
In this regime, the contact line dynamics should be comparable to those of a vesicle adhering to a flat, rigid substrate.
The contact line velocity can then be calculated as \cite{Blount2015}:
\begin{equation}
    \mathrm{v}_c=0.2\left( \frac{\omega}{\eta} \right)\left( \frac{\omega \mathrm{h}_0^2}{\kappa_\mathrm{b}} \right)^{1/4}
\end{equation}
We obtain contact line velocities $\mathrm{v}_c$ between $11$ and $20~\mathrm{\mu m/s}$.
This corresponds to wrapping velocities similar to those presented in fig. \ref{fig:Exp}B for the free particle experiments.
It should be noted that the contact line velocity is independent of particle size, matching the peak velocities of the observed wrapping events. 

In conclusion, we have characterized how wrapping dynamics change with increasing adhesion energy.
As expected, higher adhesion leads to faster wrapping, but ultimately, the dynamics of wrapping by a floppy membrane appear to be controlled by dissipation at the contact line.
As a consequence, wrapping velocities are independent of particle size, and so the time taken for an object to be wrapped is proportional to the size of the object. 
While the qualitative shape of the uptake over time is captured by a quasistatic model, differences remain.
In a quasistatic regime, the $R_\Rel$-trajectories in the free and fixed particle cases are expected to be identical.
We do however see a suppressed rebound in the latter.

Our micron-scale experiments have clear connections to the interactions of microplastics with living cells \cite{Browne2008,vonMoos2012} as well as drug delivery pathways into cells.
The incorporation of objects into a lipid membrane in the same regime (\textit{i.e.} $\ell_D\ll R_\Pa$) as our experiment are likely to follow similar dynamics to those presented here.
We expect different dynamics for objects with $\ell_D\gg R_\Pa$.
Exploring model systems in these regimes will help to establish the physical foundations for an understanding of membrane-particle interactions over a wide range of scales.

We acknowledge funding from grant number 172824 of the Swiss National Science Foundation.

\beginsupplement

\appendix

\section{Materials \& Methods}

\subsection{Materials}
1-Palmitoyl-2-oleoyl-sn-glycero-3-phosphocholine (POPC) and 1,2-dioleoyl-sn-glycero-3-phosphoethanolamine-N-(lissamine rhodamine B sulfonyl) (ammonium salt) (Rh-DOPE) were purchased from Avanti Polar Lipids, Inc. (Alabaster, Alabama).
D-(+)-glucose (BioXtra, $\geq$ 99.5\%) and sucrose (BioXtra, $\geq$99.5 \%) were purchased from Sigma Life Science. 
NaCl (ACS reagent, $\geq$99.0\%), poly(ethylene glycol) diacrylate with an average molecular weight $M_n=700$ and chloroform was bought from Sigma-Aldrich.
Ethanol, absolute was purchased from Fisher Chemicals.
3-(Trimethoxysilyl)propyl methacrylate was purchased from TCI (Tokyo Chemical Industry). 
Poly(ethylene oxide) (also called polyethylene glycol, PEG) powder with average Mv of 100,000, 2-Hydroxy-4'-(2-hydroxyethoxy)-2-methylpropiophenone (Irgacure 2959) as well as poly(ethylene glycol)-block-poly(propylene glycol)-block-poly(ethylene glycol) (PEG-PPG-PEG, Pluronic F108, average Mn $\sim$ 14,600) were bought from Aldrich Chemistry.
Fluorescent polystyrene-particles with a diameter of 1.08 $\mathrm{\mu m}$ and 2.07 $\mathrm{\mu m}$  were purchased from Microparticles GmbH (Berlin, Germany).\\

All chemicals were used as received.\\

\subsection{Electroformation of GUVs}
 POPC was used to make giant unilamellar vesicles by electroformation \cite{Angelova1986,Angelova1992}.
 Rhodamine tagged lipids (Rh-DOPE) are added in the low concentration of 1\%.
 50 $\mu L$ of a 1 mM solution of these lipids was deposited on two Platinum wires 5 mm apart using a glass syringe (Hamilton).
 The two wires are part of a PTFE chamber which is filled with a solution of 280 mOsm/kg sucrose and sealed with parafilm covered PMMA windows
 Both wires are connected electrically to a signal generator (Keysight 33210A).
 The electroformation protocol consists of gentle increase in AC-voltage over 25 minutes hour from 0 to 5 V with a fixed frequency of 10 Hz.
 After the voltage reaches 5 V it is left for two hours.
 The frequency is then lowered to 5 Hz for another 30 minutes.
 Vesicles with varying sizes between a few $\mathrm{\mu m}$ and up to 50 $\mathrm{\mu m}$ are taken out and stored in the 280 mM sucrose solution from the chamber at 4 $^{\circ}$C where they are stable for up to a few weeks.

\subsection{Experimental Details}
The samples were prepared as is described in \cite{Spanke2020}.
The substrate was covered in a thin layer of PEG-DA hydrogel preventing the GUVs from bursting as they settle.
An imaging spacer (Grace Bio-Labs SecureSeal™ imaging spacer  purchased through Sigma-Aldrich) was placed on the hydrogel-coated substrate.
The sample volume was then filled with 70 $\mathrm{\mu L}$ depletion medium, 0.3 to 0.5 $\mathrm{\mu L}$ of a 0.025 wt\% particle suspension and 10 $\mathrm{\mu L}$ sucrose solution containing the GUVs before sealing it.
 
 \subsection{Optical Microscopy and Micromanipulation}
  
   \begin{figure}
\includegraphics[width=0.9\columnwidth]{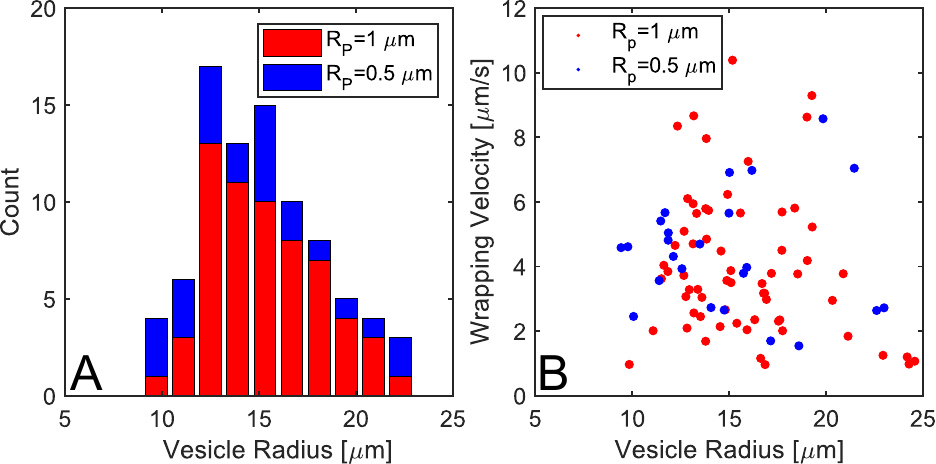}
\caption{\label{fig:S_Rv} \emph{Vesicle Size}.  
\emph{(A)} Histogram of vesicle radii $R_V$.
GUVs wrapping $1.04$ and $0.54~\mathrm{\mu m}$ radii particles are indicated in red and blue, respectively.
\emph{(B)} Wrapping velocity for each experiment versus the vesicle radius.
GUVs wrapping $1.04$ and $0.54~\mathrm{\mu m}$ radii particles are indicated in red and blue, respectively.
}
\end{figure}

 Experiments using optical tweezers were done with Nikon Ti Eclipse inverted microscopes using a 60x water immersion objective lens.
 Videos were taken at 1000 frames per second with a Hamamatsu ORCA-Flash 4.0, C13440.
 The trapping laser was a LUXX 785-200 Laser from Omicron Laserage Laserprodukte GmbH with a wavelength of a wavelength of 785 nm and a maximum power of 200 mW.
 The initially vertically polarized laser is sent through a half-wave plate and a polarizing beamsplitter.
 By correctly orienting the half-wave plate it is possible to tune the laserpower entering the sample while keeping the laser output constant.
 past the beamsplitter the still linearly polarized laserbeam is circularized by a quarter-wave plate.
 The optical trap stiffness was calibrated passively using the equipartition theorem and Boltzmann statistics method described in \cite{Sarshar2014}.
 The stiffnesses were measured to be between 1.3 and 1.7 $\mathrm{pN/\mu m}$.\\
 All vesicles used for experiments had negligible membrane tension.
 This was shown in a previous paper using the exact same system \cite{Spanke2020}.
 We defined the vesicle radius $R_\V$ as the average of major and minor axis.
 All Vesicles used for experiments were between $9.8$ and $24.6~\mathrm{\mu m}$ in radius.
 Fig. \ref{fig:S_Rv}A shows a histogram of vesicle sizes.
 We did not observe a correlation between vesicle size and wrapping velocity, as can be seen in fig. \ref{fig:S_Rv}B.\\
 
 \subsection{Quasistatic Model}

\begin{figure}
\includegraphics[width=0.9\columnwidth]{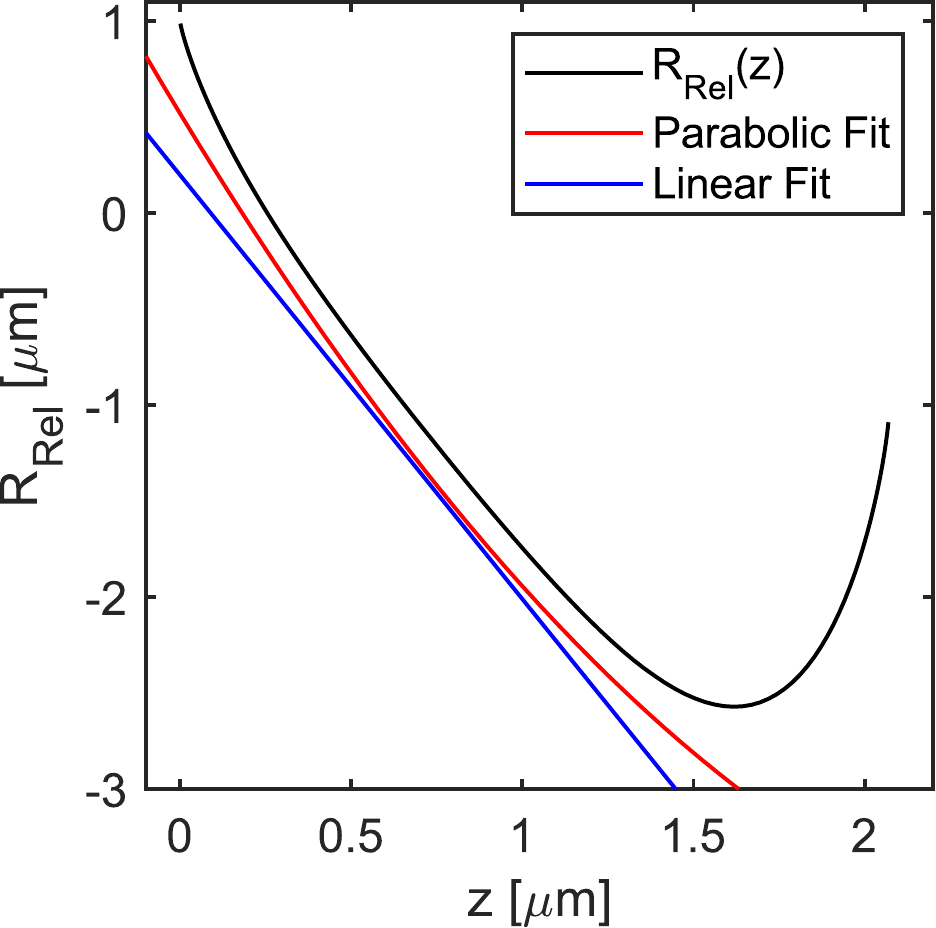}
\caption{\label{fig:S_ModelDetails} \emph{Model Details}.  
$R_\Rel(z)$ based on the quasistatic model for a particle to vesicle radius ratio of $0.08$.
This ratio corresponds to a vesicle radius $R_\V$ of $13~\mu m$, a typical size observed in experiments, and a particle radius $R_\Pa$ of $1.04~\mu m$.
A parabolic fit ($R_\Rel(z)=0.48z^2-2.95z+0.72$) is shown in red.
A linear fit ($R_\Rel(z)=-2.2z+0.50$) is shown in blue.
The parabolic and linear fits are shifted by -0.2 and -0.3 for visibility. 
}
\end{figure}

The energy of a small particle in contact with a much larger vesicle with local mean curvature $M$ and spontaneous curvature $m$ can be written as \cite{Agudo-Canalejo2017}:

\begin{equation}
E(q)=16\pi\kappa_\mathrm{b} R_\Pa [(m-M_\co)q + (M-m)q(1-q)]
\label{eq:E1}
\end{equation}
where $q=A_\mathrm{bo}/A_\Pa$ is the fraction of the particle surface bound to the membrane \cite{Agudo-Canalejo2017} and
\begin{equation}
M_\co \equiv \sqrt{\frac{\omega}{2\kappa_\mathrm{b}}}-\frac{1}{R_\Pa}
\label{eq:Mco}
\end{equation}
In this notation the condition for spontaneous wrapping is simply $M_\co = M$.
For small particles Eq. (\ref{eq:Mco}) can be rewritten as
\begin{equation}
\frac{\omega R_\Pa^2}{2\kappa_\mathrm{b}} \approx 1 + 2 M_\co R_\Pa
\end{equation}
Similarly, from the condition for spontaneous wrapping, we can write the critical adhesion $\omega_\crit$ for spontaneous wrapping as
\begin{equation}
\frac{\omega_\crit R_\Pa^2}{2\kappa_\mathrm{b}} \approx 1 + 2 M R_\Pa 
\end{equation}
The distance $z$ between the bottom of the particle and the contact line in the perpendicular direction to the undisturbed membrane is related to the degree of wrapping as  $z=2R_\Pa q$.

Making all these replacements in (\ref{eq:E1}), ignoring spontaneous curvature $m=0$ \cite{Spanke2020}, and taking $M=1/R_\V$, the vesicle curvature, we can rewrite the energy as
\begin{equation}
E(z)=-8\pi\kappa_\mathrm{b} \left[ \frac{(\omega-\omega_\crit)R_\Pa^2}{2\kappa_\mathrm{b}} \frac{z}{2R_\Pa} + 2 \frac{R_\Pa}{R_\V} \left( \frac{z}{2R_\Pa} \right)^2  \right]
\label{eq:E2}
\end{equation}

$E(z)$ has a parabolic shape as a function of $z$, and as expected the free state becomes unstable when $\omega=\omega_\crit$.

For vesicles much larger than the particle size, $R_\V \gg R_\Pa$, this simplifies to
\begin{equation}
E(z)=-2\pi R_\Pa (\omega-\omega_\crit) z
\label{eq:E3}
\end{equation}

 As discussed in the main text we asssume overdamped dynamics, \textit{i.e}.
\begin{equation}
\dot{z} = -\mu E'(z)
\label{eq:dyn1}
\end{equation}
with $\mu$ being the mobility assumed to be only a function of $R_\Pa$ and $R_\V$.
Integrating eq. (\ref{eq:dyn1}) with the full $E(z)$ shown in eq. (\ref{eq:E2}) we obtain
\begin{equation}
z(t)= \frac{(\omega-\omega_\crit)R_\Pa^2 R_\V}{4\kappa_\mathrm{b}} \left[ \exp\left( \frac{\mu 8 \pi \kappa_\mathrm{b}}{R_\Pa R_\V} t \right) -1 \right]
\label{eq:dyn2}
\end{equation}

At short times $t$ or for $R_\V \gg R_\Pa$ this equation simplifies to
\begin{equation}
z(t)\approx \mu2\pi \left( \omega - \omega_\crit \right) R_\Pa t
\label{eq:dyn2}
\end{equation}
Fig. \ref{fig:S_ModelDetails} shows $R_\Rel$ extracted from the membrane shapes obtained through bending energy minimization as a function of the wrapping degree $z$ for a particle to vesicle radius ratio of $0.08$.
This ratio corresponds to a vesicle radius $R_\V$ of $13~\mu m$, a typical size observed in experiments, and a particle radius $R_\Pa$ of $1.04~\mu m$.
A parabolic and linear fit between $20\%$ and $60\%$ of the particle diameter are shown in red and blue respectively.
The linear fit is described by $R_\Rel(z)=-2.2z+0.5$.
Combining this with eq. \ref{eq:dyn2} $R_\Rel(t)$ is given by
\begin{equation}
R_\Rel(t)=-4.4\mu \pi R_\Pa (\omega-\omega_\crit)t
\label{eq:RReloft}
\end{equation}
This equation describes the experimental $R_\Rel(t)$ best during the nearly linear part of the initial wrapping.
Taking the derivative of $R_\Rel$ with respect to time and solving for the mobility $\mu$ we arrive at:
\begin{equation}
\mu=\frac{-\dot{R}_\Rel}{4.4\pi \left( \omega - \omega_\crit \right) R_\Pa}
\label{eq:mobility2mu}
\end{equation}

With $\dot{R}_\Rel=-6.1~\mathrm{\mu m/s}$ for the average wrapping velocity at the highest adhesion energy density $\omega$ of $1.7~\mathrm{\mu J/m^2}$ and $\omega_\crit=0.65~\mathrm{\mu J/m^2}$ the mobility for a particle with a radius $R_\Pa$ of $1.04~\mathrm{\mu m}$ is calculated to be $4.2\times10^5~\mathrm{(Ns/m)}^{-1}$.
 
 \subsection{Influence of Thermal Fluctuations on Final $R_\Rel$-position}
 
 \begin{figure}
\includegraphics[width=0.9\columnwidth]{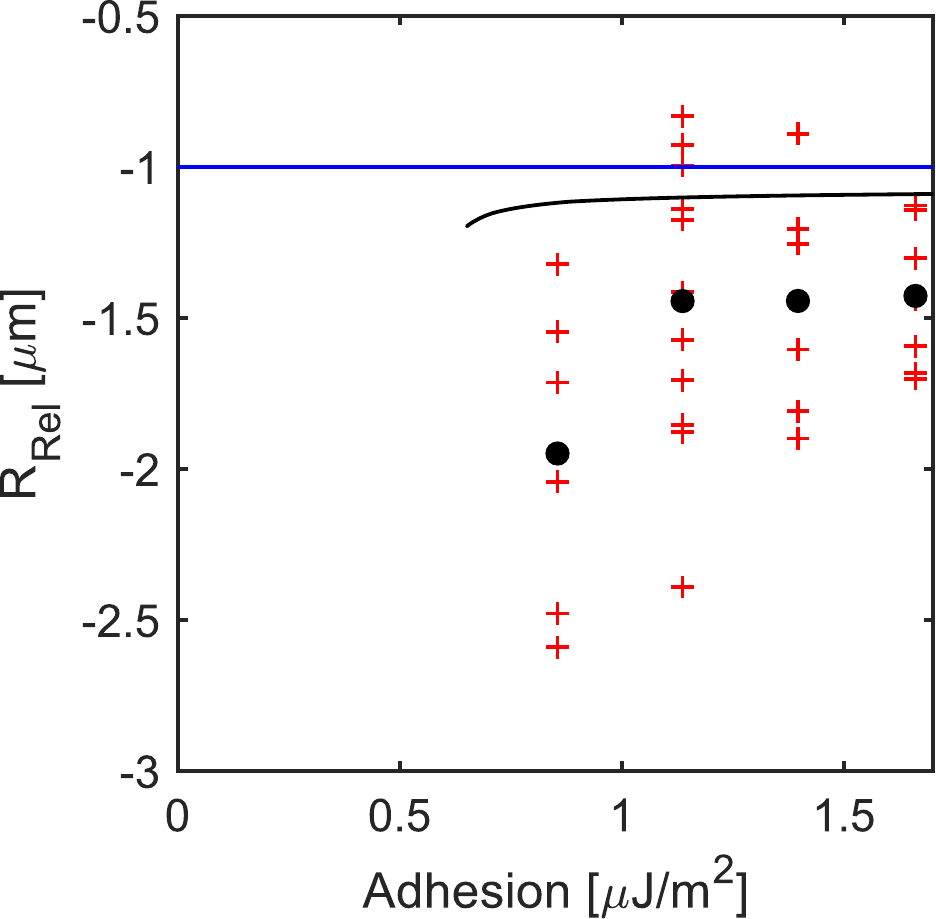}
\caption{\label{fig:S_FinalPos_Thermal} \emph{Effect of Thermal Fluctuations}.  
The final $R_\Rel$-position at which the particle settles after wrapping for a particle to vesicle radius ratio of $0.08$.
This ratio corresponds to a vesicle radius $R_\V$ of $13~\mu m$, a typical size observed in experiments, and a particle radius $R_\Pa$ of $1.04~\mu m$. The red x indicate each individual experiment and the black dot the average of those.
The blue line is the expected final position of the particle from quasistatic models.
The black line indicates the average $R_\Rel$-position considering thermal fluctuations.
}
\end{figure}
 
 It is possible to estimate the final $R_\Rel$-position of a particle taking the membrane thermal fluctuations into account using eq. \ref{eq:E2}.
 The average distance $z$, the distance $z$ between the bottom of the particle and the contact line in the perpendicular direction to the undisturbed membrane, is given by
 \begin{equation}
     \langle z \rangle = \frac{\int_{0}^{2R_\Pa} ze^{-E(z)/k_BT} \,dz}{\int_{0}^{2R_\Pa} e^{-E(z)/k_BT} \,dz}
 \end{equation}
 At the critical adhesion energy density $\omega=\omega_\crit$, $\kappa_\mathrm{b}=33~\mathrm{k_B T}$ and $R_P/R_V=0.08$ we obtain $\langle z \rangle=1.992 R_\Pa$.
 Due to the high slope of $R_\Rel(z)$ for $z$ close to $2R_\Pa$ this corresponds to $\langle R_\Rel \rangle =-1.197~\mathrm{\mu m}$.
 Fig. \ref{fig:S_FinalPos_Thermal} shows the final positions of a particle as observed in experiments for a particle to vesicle radius ratio of $0.08$.
This ratio corresponds to a vesicle radius $R_\V$ of $13~\mu m$, a typical size observed in experiments, and a particle radius $R_\Pa$ of $1.04~\mu m$.
The black line shows $\langle R_\Rel \rangle$ as a function of the adhesion energy density $\omega$.
 
 \subsection{Spontaneous Wrapping of Particles with a Radius of $0.54~\mathrm{\mu m}$}
 
\begin{figure}
\includegraphics[width=0.9\columnwidth]{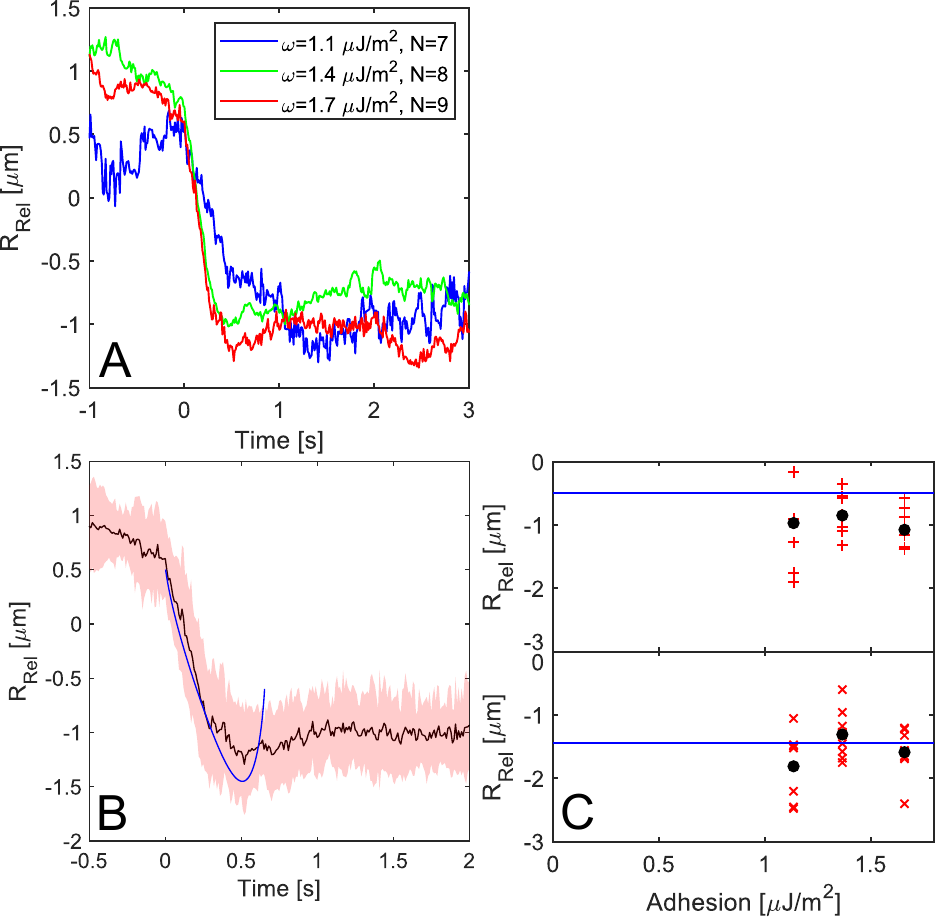}
\caption{\label{fig:S_1mu} \emph{Effect of Particle Size}.  
\emph{(A)} Averaged $R_\Rel$-curves for a particle with a radius of $0.54~\mathrm{\mu m}$ with increasing adhesion energy densities. Each curve is the average of \textit{N} individual experiments, as is indicated in the legend. 
\emph{(B)} Averaged $R_\Rel$-curve for a particle with a radius of $0.54~\mathrm{\mu m}$ at an adhesion energy density of $1.7~\mathrm{\mu J/m^2}$.
The standard deviation in $R_\Rel$ at each timepoint is indicated in red.
The blue line is the modelled $R_\Rel$.
\emph{(C)} The final $R_\Rel$-position at which the particle settles after wrapping and the deepest indentation during the process. The red x indicate each individual experiment and the black dot the average of those.
The blue line is the expected final position and deepest indentation of the particle from quasistatic models.
}
\end{figure}

Fig. \ref{fig:S_1mu}A shows $R_\Rel$ over time for three different adhesion energy densities ranging from $1.1$ to $1.7~\mathrm{\mu J/m^2}$.
Each curve is the average of \textit{N} individual experiments, as is indicated in the legend.
Before averaging, the individual curves were median-filtered over ten frames to filter out noise.\\
A modelled $R_\Rel$ curve for a particle with a radius of $0.54~\mathrm{\mu m}$ at an adhesion energy density of $1.7~\mathrm{\mu J/m^2}$ is shown in Fig. \ref{fig:Model}B in blue.
The average $R_\Rel$-curve from experiments is shown in black.
An error for each time point is indicated by the area shaded in red.
A linear fit to the modelled $R_\Rel(z)$ between $20\%$ and $60\%$ of the particle diameter gives a slope of $-2.4$.
The mobility is then given by
\begin{equation}
\mu=\frac{-\dot{R}_\Rel}{4.8\pi \left( \omega - \omega_\crit \right) R_\Pa}
\label{eq:mobility1mu}
\end{equation}
and equals $6.7\times10^{5}~\mathrm{(Ns/m)}^{-1}$, about $1.6$ times the mobility for the $1.04~\mathrm{\mu m}$ particles.
As discussed in the main text the depth at which the particle settles after the wrapping process will be at $R_\Rel=-R_\Pa$ and the rebound depth, or the deepest indentation of the particle into the membrane will be at $R_\Rel\approx-2.5R_\Pa$.
These two $R_\Rel$ coordinates will not change with changing adhesion energy density.
In Experiments we still consistently see lower final depths although the deepest indentation is observed to match with the quasistatic model as is visible in fig. \ref{fig:S_1mu}C.

 \subsection{Supplementary Videos}
 
 \begin{itemize}
    \item Movie S1: free PS particle with a radius of $1.04~\mathrm{\mu m}$ being wrapped by a GUV with $R_V\approx13.0\ \mathrm{\mu m}$.
    The sample contains 0.67 wt\% PEG with $M_w\approx100'000$ g/mol, corresponding to an adhesion energy density of $1.7~\mathrm{\mu J/m^2}$.
    The video was acquired at 1000 frames per second and is replayed at 100 frames per second.
    
    \item Movie S2: trapped PS particle with a radius of $1.04~\mathrm{\mu m}$ being wrapped by a GUV with $R_V\approx14.5\ \mathrm{\mu m}$.
    The sample contains 0.67 wt\% PEG with $M_w\approx100'000$ g/mol, corresponding to an adhesion energy density of $1.7~\mathrm{\mu J/m^2}$.
    The video was acquired at 1000 frames per second and is replayed at 100 frames per second.
 \end{itemize}
 
 \subsection{Supplementary Data}
\emph{ExpData.xlsx:}
Excel array with the following columns:
\begin{itemize}
\item Column 1: Particle radius in meters
\item Column 2: wt\% of PEG100K in the sample
\item Column 3: adhesion energy density in $\mathrm{J/m^2}$
\item Column 4: 1/2 of the major axis of the GUV in meters
\item Column 5: 1/2 of the minor axis of the GUV in meters
\item Column 6: Peak force experienced by the particle. Given as $NaN$ for free particle experiments.
\item Column 7: Wrapping velocity in $\mathrm{m/s}$
\item Column 8: Type of experiment. 0 - free particle, 1 - trapped particle.
\end{itemize} 


\end{document}